\documentclass[conference]{IEEEtran}

\usepackage{cite}
\usepackage{amsmath,amssymb,amsfonts}
\usepackage{algorithmic}
\usepackage{algorithm}
\usepackage{graphicx}
\usepackage{textcomp}
\usepackage{xcolor}
\usepackage{url}
\usepackage{subcaption}
\usepackage{adjustbox}
\usepackage{listings}
\begin{document}
\title{Hybrid Materialization in a Disk-Based Column-Store}
\date{}

\author{
\IEEEauthorblockN{Evgeniy Klyuchikov, Elena Mikhailova, George Chernishev}
\IEEEauthorblockA{Saint Petersburg State University \\Saint-Petersburg, Russia\\ evgeniy.klyuchikov@gmail.com, egmichailova@mail.ru, chernishev@gmail.com\\}}
\maketitle

\begin{abstract}

In column-oriented query processing, a materialization strategy determines when lightweight positions (row IDs) are translated into tuples. It is an important part of column-store architecture, since it defines the class of supported query plans, and, therefore, impacts the overall system performance.

In this paper we continue investigating materialization strategies for a distributed disk-based column-store. We start with demonstrating cases when existing approaches impose fundamental limitations on the resulting system performance. Then, in order to address them, we propose a new hybrid materialization model. The main feature of hybrid materialization is the ability to manipulate  both positions and values at the same time. This way, query engine can flexibly combine advantages of all the existing strategies and support a new class of query plans. Moreover, hybrid materialization allows the query engine to flexibly customize the materialization policy of individual attributes. 

We describe our vision of how hybrid materialization can be implemented in a columnar system. As an example, we use PosDB~--- a distributed, disk-based column-store. We present necessary data structures, the internals of a hybrid operator, and describe the algebra of such operators. Based on this implementation, we evaluate performance of late, ultra-late, and hybrid materialization strategies in several scenarios based on TPC-H queries. Our experiments demonstrate that hybrid materialization is almost two times faster than its counterparts, while providing a more flexible query model.

\end{abstract}

\section{Introduction}

Column-stores are well-known for their outstanding performance on analytic workloads~\cite{cstore_materialization, DOLAP22-newarch}. One of the main factors of their success is the ability to work with individual columns, thus reading data from disk only when it is absolutely necessary. Apart from that, such processing allows the query executor to benefit from compression and vectorized computations~\cite{cstore_compression, vectorwise, MEDI21-compression}.

While working with individual attribute values, it is still necessary to reference a subset of table tuples from which the values originate. For this purpose, column-stores usually use row numbers, named positions, or row ids~\cite{Abadi:2013:DIM:2602024}. Using positions is convenient for the query engine, but they ultimately will have to be transformed into attribute values and stitched together to form resulting tuples.

However, most modern column-stores significantly differ from the vision that was held by the pioneers~\cite{monet_2012, Boncz:1999:MPQ:765509.765511, c-store}. Research prototypes of the 00's featured fluent data position handling, enabling the query engine to explicitly use positions during query execution. This made available a number of optimizations and techniques that offered various benefits for query processing. Unfortunately, the majority of mainstream systems did not follow the original ideas and instead resorted to the simpler ones (thus gaining the name of naive column-stores~\cite{Abadi:2013:DIM:2602024}). Therefore, we differentiate the ``position-enabled'' column-stores from the rest as the column-stores that are able to reap benefits from explicit position manipulation inside their engine.

The process of transitioning from positions to values in a position-enabled column-store is called materialization. Thus, a materialization strategy determines when and how values are materialized in the query plan. The materialization strategy is an essential part of a position-enabled column-store architecture and can change system performance significantly.

In our previous work~\cite{DOLAP22-newarch} we provided a comprehensive study of basic principles that various systems use to materialize data. We compared existing early and late materialization strategies, and proposed a new, ultra-late materialization to process analytic workloads even faster.

\begin{figure}[htb]
    \centering
    \includegraphics[width=1\linewidth]{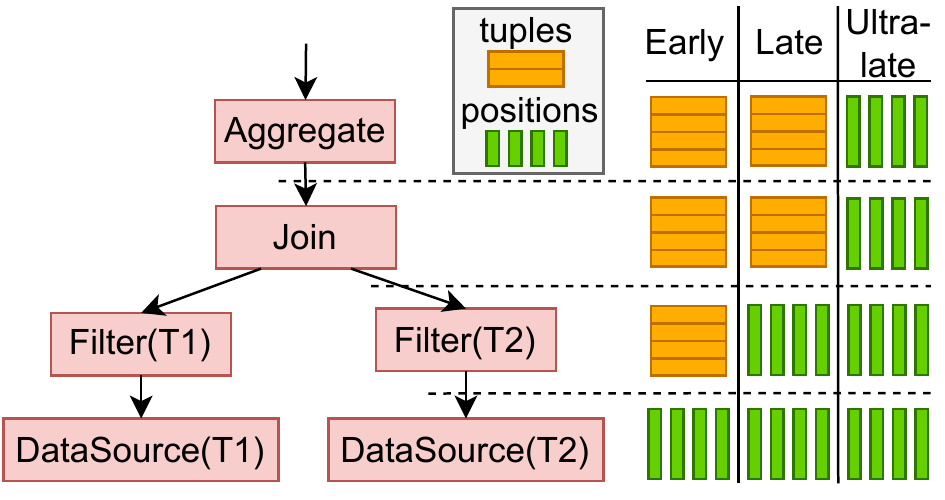}
    \caption{SPJ query with materialization points}
    \label{fig:mat_points}
\end{figure}

To get an overall idea of various materialization strategies, consider Fig.~\ref{fig:mat_points}. It shows an SPJ query plan in a column-store while simultaneously describing data representation evolution over different query stages. The early, late, and ultra-late strategies use positions at the bottom of the query plan, and at some \textit{materialization point} they transform positions to attribute values. After that, query execution continues with tuples, similarly to classic row-stores.

\begin{table*}[!ht]
\centering
\caption{Aspects of materialization strategies}
\begin{tabular}{llllll}

Materialization & \begin{tabular}[c]{@{}l@{}}Fast predicates\end{tabular} & \begin{tabular}[c]{@{}l@{}} Re-read during predicates \end{tabular} & \begin{tabular}[c]{@{}l@{}}Pre-read before joins\end{tabular} & Re-read during joins & Out-of-order probing\\ \hline \hline

Early & No & No & Yes & No & No \\ \hline
Late  & Yes & Yes & Yes & No & No \\ \hline
Ultra-late & Yes & Yes & No & Yes & Yes \\ \hline
Hybrid & ? & ? & ? & ? & ?
\end{tabular}
\label{tbl:m_compare}
\end{table*}

Early materialization constructs tuples before filters, late before joins, and ultra-late before aggregation. Our investigation has demonstrated that ultra-late materialization can benefit from reading attribute data after filtering joins, which leads to a more than two-time speedup~\cite{DOLAP22-newarch} on workloads such as the Star Schema Benchmark~\cite{SSB}.

However, we believe that all the three strategies are limited by their design. They use either positions or tuples at any given moment of time (but not both), which prevents flexible materialization of individual attributes. Thus, in this paper we present a new, hybrid materialization strategy, which is free from the drawbacks of its counterparts.

The idea of hybrid materialization is to pass both positions and values through operators. This way, we can combine advantages of all other strategies. Similarly to ultra-late materialization, we can read some attributes of a query just before aggregation. However, if for some attributes such reads are expensive~\cite{vertica}, we can flexibly cache these specific attributes much earlier and pass them together with positions.

In this paper, we will investigate hybrid materialization in disk-based, position-enabled column-stores. We will analyze the early, late, and ultra-late strategies to show cases where they are fundamentally limited due to one-step materialization. We will present our vision of how hybrid materialization can be implemented in a disk-based column-store such as PosDB\footnote{https://pos-db.com/}. We are going to describe hybrid data structures, internals of a hybrid operator, and the overall hybrid query model. Finally, we will demonstrate query plans with hybrid materialization and will experimentally evaluate its performance compared to late and ultra-late materialization to prove our findings.

The overall contribution of this paper is the following:
\begin{enumerate}
    \item A qualitative analysis of early, late, and ultra-late materialization strategies. We highlight cases in which they can be improved via hybrid position-value processing.
    \item A hybrid query model. We describe hybrid data structures, hybrid operator internals, and algebra of hybrid operators. We also provide insights on how to safely introduce the hybrid strategy into a system with either early, late, or ultra-late materialization.    
    \item An experimental evaluation of hybrid materialization compared to the late and ultra-late, using modified queries from the TPC-H benchmark~\cite{TPCH}.
\end{enumerate}

This paper is organized as follows. In Section~\ref{sec:background} we overview the basics of existing materialization strategies, show where hybrid materialization can be beneficial, and discuss related work. Section~\ref{sec:model} shows how hybrid materialization can be implemented in a disk-based column-store based on our experience in PosDB. In Section~\ref{sec:advantages} we discuss advantages of hybrid materialization in local and distributed cases in more detail. Then, Section~\ref{sec:experiments} contains experimental evaluation of various materialization strategies. Finally, in Sections~\ref{sec:future} and~\ref{sec:conclusion} we present our future plans and conclude the paper.

\section{Background and Related Work}
\label{sec:background}

\subsection{Background}

To provide a more detailed motivation for hybrid materialization and outline the relevant related studies, we are going to first overview the way column-stores materialize data. For this, we shortly review the main aspects of existing materialization strategies and show where they are limited due to one-step materialization.

We use the SPJ class of queries for our analysis and experiments. Such queries are simple enough, and at the same time contain operators whose performance depends significantly on the chosen materialization strategy. Namely, we concentrate on joins and filters (predicates).

As shown in Fig.~\ref{fig:mat_points}, early, late, and ultra-late materialization strategies differ in the moment of time when positions are transformed into tuples. Thus, they differ in how operators process data in memory and how this data is loaded from disk.

Table~\ref{tbl:m_compare} summarizes five important aspects of data processing in various materialization strategies.

Early materialization uses columnar storage to read only the necessary attributes and then switches to tuple processing. With this strategy, a column-store is similar to a regular row-store.

Late materialization uses positions further through the plan to make predicates fast via processing homogeneous columnar data loaded on-demand. It gives a significant performance boost, but some attributes can be read multiple times.

Ultra-late materialization pushes positions yet further to speed up joins, similarly to predicates. Operating on positions, joins can have smaller hash tables and load attributes only when necessary. The latter can be especially efficient in workflows with ``filtering'' joins that leave only a small part of records~\cite{DOLAP22-newarch}.

Two known drawbacks of ultra-late materialization is that it is vulnerable to re-reads during joins, and that it can also be forced to read attribute data by an unordered list of positions. The latter requires expensive random disk access and is known as the \textit{out-of-order probing problem}~\cite{columns_tutorial}. 

Hybrid materialization may be a compromise between all the existing materialization strategies. Its core idea is to pass both positions and values at some parts of a query plan. Therefore, the query engine can benefit from using positions and at the same time, early materialize those attributes that are expensive to reaccess. This way, hybrid materialization allows the query engine to customize each of the five aspects for each individual attribute or table. As the result, it provides the best possible strategy for a given query.

\subsection{Related Work}

In this section, we briefly discuss existing papers that concern materialization strategies or hybrid position-value processing.

Materialization strategy as a term was first introduced by the authors of C-Store~\cite{cstore_materialization}, a disk-based column-store, to discuss how data is translated from columns on disk to resulting tuples in memory. They described the idea of materializing some data during query processing and explained the related costs. However, they mainly focused on early and late materialization and their performance.

Interestingly, the authors present several operators that read attribute data from disk and one of them actually uses both positions and values at the same time~\cite{cstore_materialization}. This means that C-Store actually supported hybrid materialization, but only within one specific operator. They also mentioned that the hybrid approach could be useful in join operators~\cite{Abadi:2013:DIM:2602024}, when the left subtree uses positions, but the right subtree uses tuples. However, they did not continue the investigation of hybrid materialization and did not discuss its usage for the entire query in the local and distributed cases.

The authos of the MonetDB system also explored various materialization strategies~\cite{Boncz:1999:MPQ:765509.765511, monet_2012}. Its operators use Binary Association Tables, special tables that consist of one or two columns. These columns contain either positions or values, and the query engine can materialize data using them on-demand. Such a specific representation was implemented because MonetDB was designed as an in-memory system, and all its operators were optimized for fast in-memory processing. It also supports only local query processing. Thus, MonetDB ideas can be only partially used for hybrid materialization in a distributed disk-based column-store.

Another series of works on hybrid position-value processing originated from PAX~\cite{pax}. PAX was an attempt to speed up a classic row-store engine by changing the way data is organized inside data pages on disk and in memory. Its core idea was to replace array of tuples within a data page to a series of minipages, one per attribute. Thus, the data of each attribute is stored separately, just like in column-stores, and all the minipages contain data for the same range of table tuples, i.e. have some virtual row ids. Such hybrid storage could be used by the query engine to process data using both positions and values up to some point.

A number of systems were built based on the PAX model. The next generation of MonetDB, Vectorwise, uses PAX to support workloads in which the total size of tables exceeds the RAM limit~\cite{vectorwise}. However, the hybrid format is used only on disk, while the Vectorwise query engine is more focused on vectorized computations, concurrent query processing, and data caching. 

As a more recent attempt, HyPer, a high-performance main-memory database for hybrid OLAP \& OLTP workloads, introduced DataBlocks~\cite{datablocks}. DataBlocks extends PAX ideas in the following way. First, each page contains a number of byte-addressable compressed chunks of attribute data. This way, they are similar to compressed columns in a column-store system. Secondly, HyPer introduces Positional Small Materialized Aggregates (PSAMs), a special type of light-weight index stored within each page. PSAM is a table of entries that for each attribute range $[b, e)$ specifies potentially matching tuples, i.e. positions in the corresponding compressed chunks. HyPer leverages such hybrid storage to significantly speed up scans (predicates on attributes). While DataBlocks ideas are interesting and similar in spirit to hybrid materialization, HyPer is a main-memory database and is designed for hybrid workloads. Thus, it differs from our case of a distributed disk-based column-store.

Finally, there was an attempt to speed up PostgreSQL for modern SSDs using PAX~\cite{FlashJoin}. The authors present a set of specialized operators, one of which, FlashJoin, uses techniques very similar to hybrid materialization. Inside a chain of FlashJoins, both positions and values can co-exist for efficient data access. In the last FlashJoin, regular tuples are reconstructed and query processing continues in a classic row-store way. 

As we can see, there is lack of investigations on hybrid materialization in disk-based column-store systems. In our previous work~\cite{DOLAP22-newarch}, hybrid materialization was shortly mentioned as a very desired optimization for distributed queries, but was not extensively studied. This paper continues with a detailed look at hybrid materialization and the scenarios in which it surpasses other strategies.

\section{Hybrid query model}
\label{sec:model}
To support a new materialization strategy, a column-store needs to provide a new data representation and a number of operators. In the case of hybrid materialization, the main difficulty is constructing a model that is flexible enough while still allowing an efficient implementation.

We did not find a detailed discussion of the hybrid query model for column-stores in research papers. Thus, we present our vision based on the implementation of hybrid materialization in PosDB.

\subsection{Data representation}

\begin{figure}[ht]
\centering
\begin{subfigure}{.098\textwidth}
    \centering
    \includegraphics{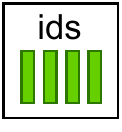}
    \caption{Positional}
    \label{fig:blocks:pos}
\end{subfigure}%
\begin{subfigure}{.135\textwidth}
    \centering
    \includegraphics{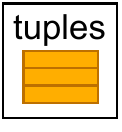}
    \caption{Tuple}
    \label{fig:blocks:tuple}
\end{subfigure}%
\begin{subfigure}{.242\textwidth}
    \centering
    \includegraphics{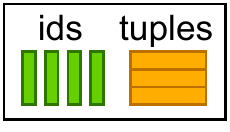}
    \caption{Hybrid}
    \label{fig:blocks:hybrid}
\end{subfigure}
\caption{Block structures in PosDB'23}
\label{fig:blocks}
\end{figure}

PosDB uses the block-based Volcano query model~\cite{graefe_query_1993}, in which each query plan consists of operators that exchange data blocks. Thus, we need a new hybrid data block that can store both positions and values at the same time, see Fig.~\ref{fig:blocks:hybrid}.

A hybrid block consists of a set of position columns and an array of tuples (both of which can be empty). A set of positional columns is actually a positional block, used by early, late and ultra-late materialization before a materialization point, see Fig.~\ref{fig:blocks:pos}. It makes column-specific operations, such as reading attribute values by positions and filtering them by predicate~\cite{Zukowski2008-ej, Abadi:2013:DIM:2602024}, efficient.

An array of tuples is taken from a tuple block, used by aforementioned materialization strategies after the materialization point, see Fig.~\ref{fig:blocks:tuple}. Tuple representation is better suited for value-intensive actions as operators can benefit from space locality in this case~\cite{Zukowski2008-ej}.

\subsection{Hybrid operator internals}
\label{sec:hybrid:internals}

Late and ultra-late materialization use a special materialization point to transform positional blocks into tuple blocks. Hybrid materialization relies on a different approach when attributes are materialized on-demand in many operators.

This means that hybrid operators should both implement their primary algorithms and deal with data transformation. We consider three main cases:

\begin{enumerate}
    \item Materializing a new attribute before an operator.
    \item Passing some input positions and values through an operator.
    \item Adding operator results to the output hybrid blocks as a column of positions or an additional attribute in tuples.
\end{enumerate}

It is possible to implement both the core operator algorithm and all the data transformations in one entity, but that would have made query plan analysis and optimization difficult. Instead, we use the FlashJoin~\cite{FlashJoin} idea of breaking each hybrid operator into three parts: \lstinline{Fetch}, \lstinline{Core}, and \lstinline{Combine}, implemented as separate physical operators.

\begin{figure}[htb]
    \centering
    \includegraphics[width=\linewidth]{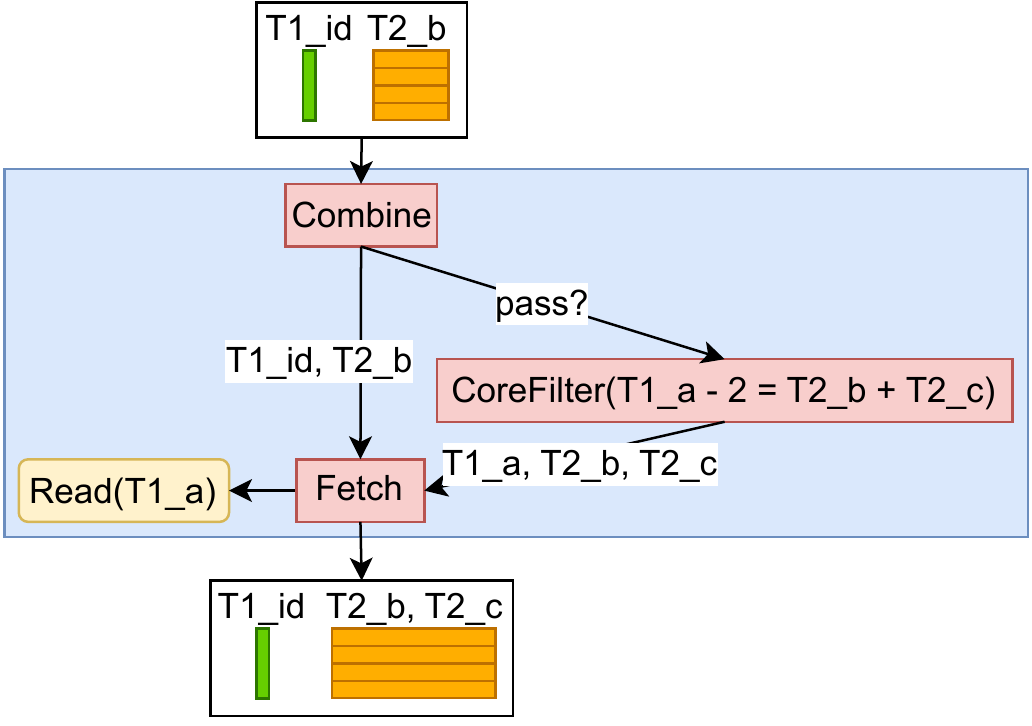}
    \caption{Internals of the hybrid self-join operator}
    \label{fig:hyb:filter}
\end{figure}

Fig.~\ref{fig:hyb:filter} shows how an example hybrid self-join is organized on the physical level. For convenience, we denote the positions of $T1$ as the $T1\_id$ column, and use similar notation further on. \lstinline{Fetch} materializes $T1.a$ and provides access to $T2.b$ and $T2.c$. Then \lstinline{CoreFilter} checks whether $T1.a - 2 = T2.b + T2.c$, and if so, \lstinline{Combine} adds positions of $T1$ and $T2.b$ to the resulting hybrid block.

\begin{figure*}[!t]
\centering
\begin{subfigure}{0.345\linewidth}%
    \centering
    \includegraphics[scale=0.95]{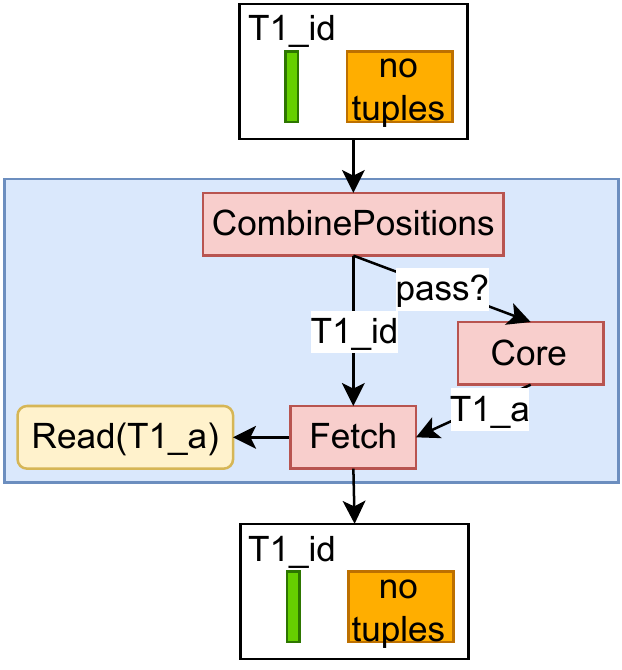}
    \caption{For positions}
    \label{fig:fetch:pos}
\end{subfigure}%
\begin{subfigure}{0.276\linewidth}%
    \centering
    \includegraphics[scale=0.95]{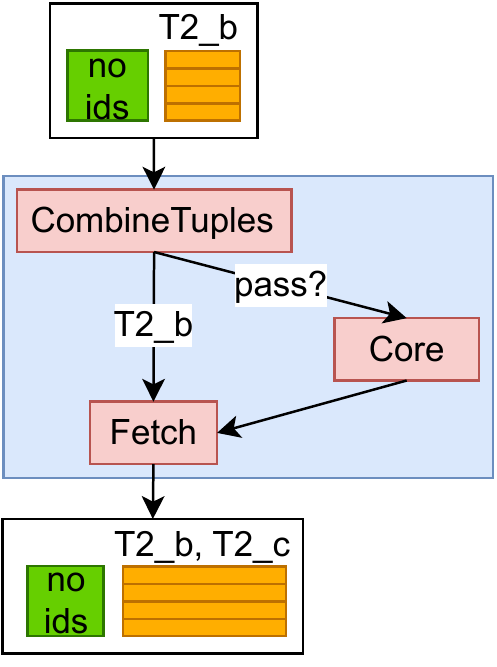}
    \caption{For tuples}
    \label{fig:fetch:tuple}
\end{subfigure}%
\begin{subfigure}{0.3789\linewidth}%
    \centering
    \includegraphics[scale=0.95]{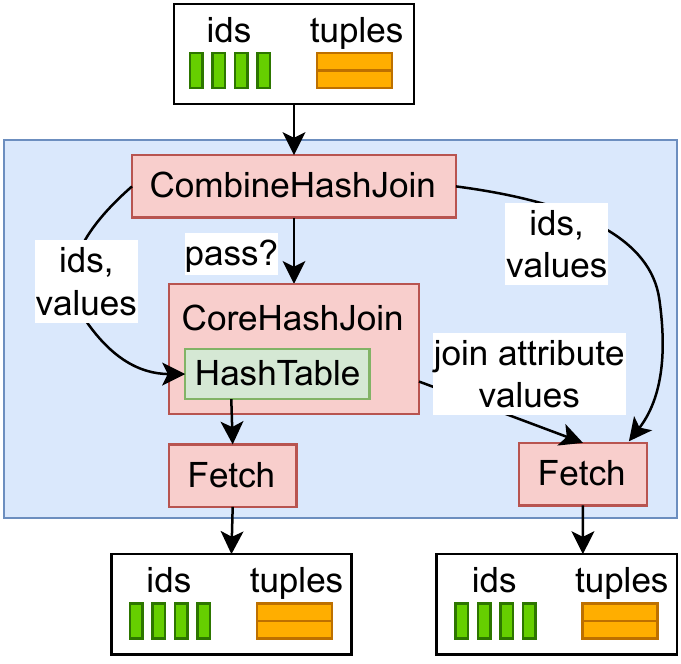}
    \caption{For HashJoin}
    \label{fig:fetch:hash}
\end{subfigure}

\caption{Fetch-Combine specializations}
\label{fig:fetch}
\end{figure*}

Note that the \lstinline{Core} part can work with any \lstinline{Fetch} implementation as long as it provides required values. Similarly, \lstinline{Core} produces data independently of how \lstinline{Combine} is implemented. Thus, we can implement any number of specialized \lstinline{Fetch} and \lstinline{Combine} operators, selecting most efficient ones during query optimization and physical plan creation.

As an example, Fig.~\ref{fig:fetch} shows three specialized \lstinline{Fetch-Combine} pairs. When we know that only positions or tuples are used, we can simplify block processing, using a classic column or row model (Fig.~\ref{fig:fetch:pos}, Fig.~\ref{fig:fetch:tuple}). This way hybrid materialization can efficiently emulate late or ultra-late materialization. Preliminary benchmarks showed a several percent difference between the emulated and ``original'' versions of materialization strategies (which is most probably due to implementation-specific aspects).

As a more interesting example, we can create a special \lstinline{CombineHashJoin} to build hybrid blocks during hybrid hash-join, as shown in Fig.~\ref{fig:fetch:hash}. \lstinline{CoreHashJoin} uses \lstinline{Fetch} of the left subtree to build a hashtable in memory. Then it sequentially processes values from \lstinline{Fetch} of the right subtree and calls \lstinline{CombineHashJoin} to build resulting blocks. \lstinline{CombineHashJoin} knows about hash-join specifics and uses data both from right \lstinline{Fetch} and hashtable inside join. Moreover, we can devise a special \lstinline{CombineHashJoin} for cases when only data from the hashtable or the right \lstinline{Fetch} is required.

Depending on the system architecture, we can provide dozens of \lstinline{Fetch-Combine} specializations, as they leave core operators intact. All we need is to enhance the query planner with heuristics to find the best specialization.

\subsection{Algebra of hybrid operators}

We have developed a minimum set of hybrid operators that is sufficient to process simple SPJ queries. Each of these operators accepts blocks of a particular type as input (possibly several arguments) and returns blocks of some other type. In the following list, types of input blocks are put before $\rightarrow$ and resulting blocks are given after. Overall, the following operators were implemented:

\begin{enumerate}
    \item HYDataSource: Nil $\rightarrow$ HybridBlocks,
    \item HYFilter, HYProject, HYMaterialize: HybridBlocks $\rightarrow$ HybridBlocks;
    \item HYHashJoin, HYNestedLoopJoin: \{HybridBlocks1, HybridBlocks2\} $\rightarrow$ HybridBlocks;
    \item HYToTuple: HybridBlocks $\rightarrow$ TupleBlocks.
\end{enumerate}

Here HYDataSource produces positions for a table, then HYMaterialize, HYProject and HYFilter add, filter and remove attribute values on-demand. HYHashJoin and HYNestedLoopJoin join two tables, and HYToTuple transform hybrid blocks to tuple blocks once all the data was materialized.

We intentionally omitted aggregate operators as we concentrate on SPJ queries without complex aggregation. Simple selections can be supported by HYMaterialize and HYProject, or by aggregate operators from the ultra-late materialization model as shown in the next section.

Each HYOperator consists of fetch, combine, and core parts. For the sake of simplicity, we omit fetch and combine details, suggesting that in each query plan they are specialized in an efficient way. We also use both HYOperator notation and detailed fetch-combine-core notation depending on the required level of detail.

\subsection{Transitioning to and from the hybrid model}

When using early, late, and ultra-late materialization, we are stuck with either positions or tuples at any given moment of time. With hybrid materialization, we can flexibly combine all the materialization models, reading values on-demand and caching them for as long as we want.

\begin{figure}[!ht]
\begin{subfigure}[b]{0.6\linewidth}\quad
    \centering
    \includegraphics[width=\linewidth]{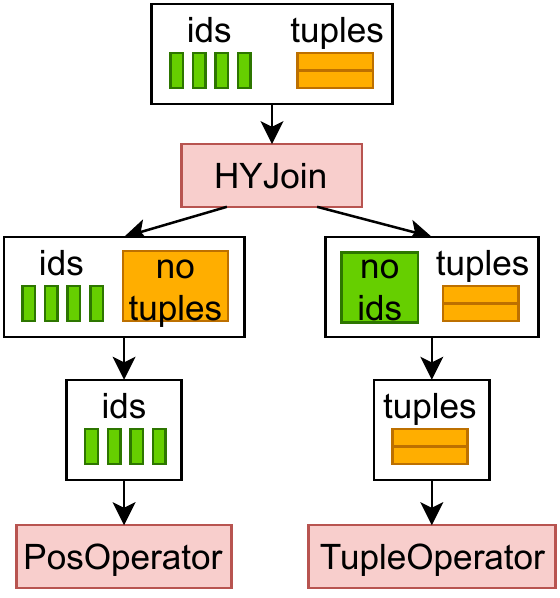}
    \caption{To hybrid}
    \label{fig:to_hybrid}
\end{subfigure}\hfill
\begin{subfigure}[b]{0.285\linewidth}%
    \centering
    \includegraphics[width=\linewidth]{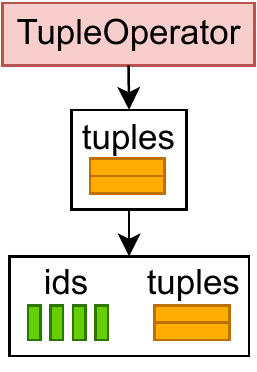}
    \caption{From hybrid}
    \label{fig:from_hybrid}
\end{subfigure}
\caption{Transition between materialization strategies}
\label{fig:hybrid_cast}
\end{figure}

For a example, we can have a join between left tree that uses ultra-late materialization and right tree that uses early materialization, see Fig.~\ref{fig:to_hybrid}. Left tree provides us a set of position columns that we treat as a hybrid block with no tuples. Right tree provides us a set of tuples that we treat as a hybrid block with no positions. Then, we can make a regular hybrid join and continue execution in a hybrid model.

Similarly, we can easily switch from hybrid module to classic tuple-based processing once all the data is materialized, as shown in Fig.~\ref{fig:from_hybrid}.

Such transitions can be useful if we have a highly optimized column or row processing engine and would like to speed up some parts of a query plan with hybrid materialization. It can be also useful to use late and ultra-late materialization when a small/some part of data is received from an external source in a tuple form.

\subsection{Example query plans}

\begin{figure}[!ht]
    \centering
    \includegraphics[width=\linewidth]{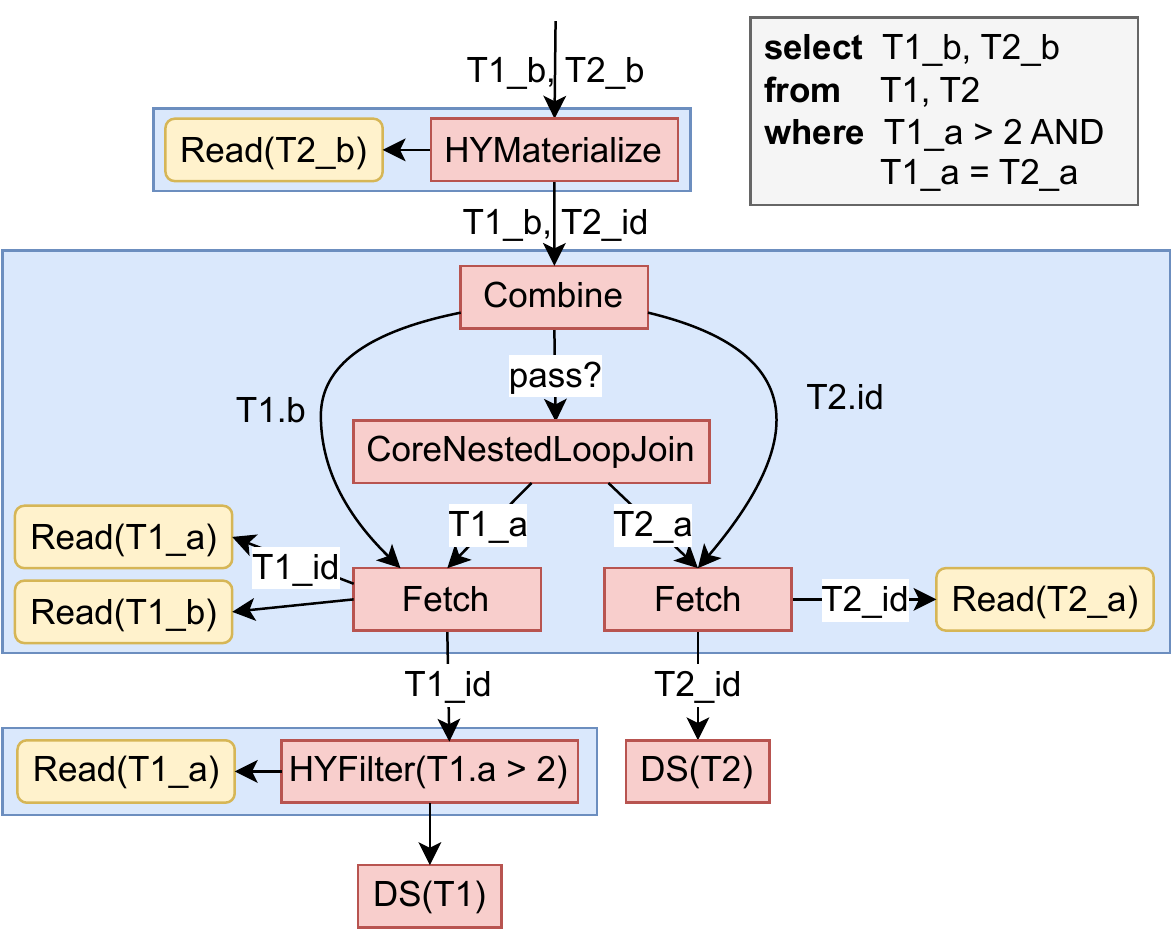}
    \caption{Simple query in a hybrid query model}
    \label{fig:hyb:example}
\end{figure}

To summarize the hybrid model discussion, let us show a simple query executed using hybrid materialization in Fig.~\ref{fig:hyb:example}.

First, HYFilter applies predicate using positions and readers as in late materialization. Then, two \lstinline{Fetch} operators receive data for $T1\_a$, $T1\_b$, and $T2\_c$, fully materializing $T1$. Hybrid nested-loop join combines $T1$ and $T2$ in hybrid blocks containing $T1\_b$ and $T2\_id$. After the join, HYMaterialize materializes $T2\_b$ using $T2\_id$ and send resulting tuples to user.

Having such hybrid plans, the question is when and why can they be more efficient than plans with late or ultra-late materialization.

\section{Hybrid materialization advantages}
\label{sec:advantages}
A classic way to compare several algorithms is to provide a theoretical model, find out interesting cases, and prove them experimentally. Building a full-fledged query model for materialization strategies requires quite a lot of space, let's use an informal analysis based on their drawbacks instead:

\begin{enumerate}

    \item Late materialization: 
    \begin{itemize}
        \item read extra data that is later discarded in subsequent operators (e.g. joins),
        \item push the data through all these operators.
    \end{itemize}
    \item Ultra-late materialization:
    \begin{itemize}
        \item read attribute values multiple times (repeated access),
        \item read attribute values from disk with random access (a so-called out-of-order probing problem).
    \end{itemize}
\end{enumerate}

Based on these drawbacks we will analyze local and distributed cases to determine when hybrid materialization can perform better.

\subsection{Local case. Ultra-late vs hybrid materialization}

Previous work~\cite{DOLAP22-newarch} has proved ultra-late materialization to be quite effective for analytic workloads. In most cases, the Volcano model and a contemporary buffer manager make repeated access to attribute data cheap. However, there are cases when ultra-late materialization forces multiple re-reads from disk or induces out-of-order probing.

\begin{figure*}[!ht]
\centering
\begin{subfigure}{0.5035\linewidth}\quad
    \centering
    \includegraphics[width=0.97\linewidth]{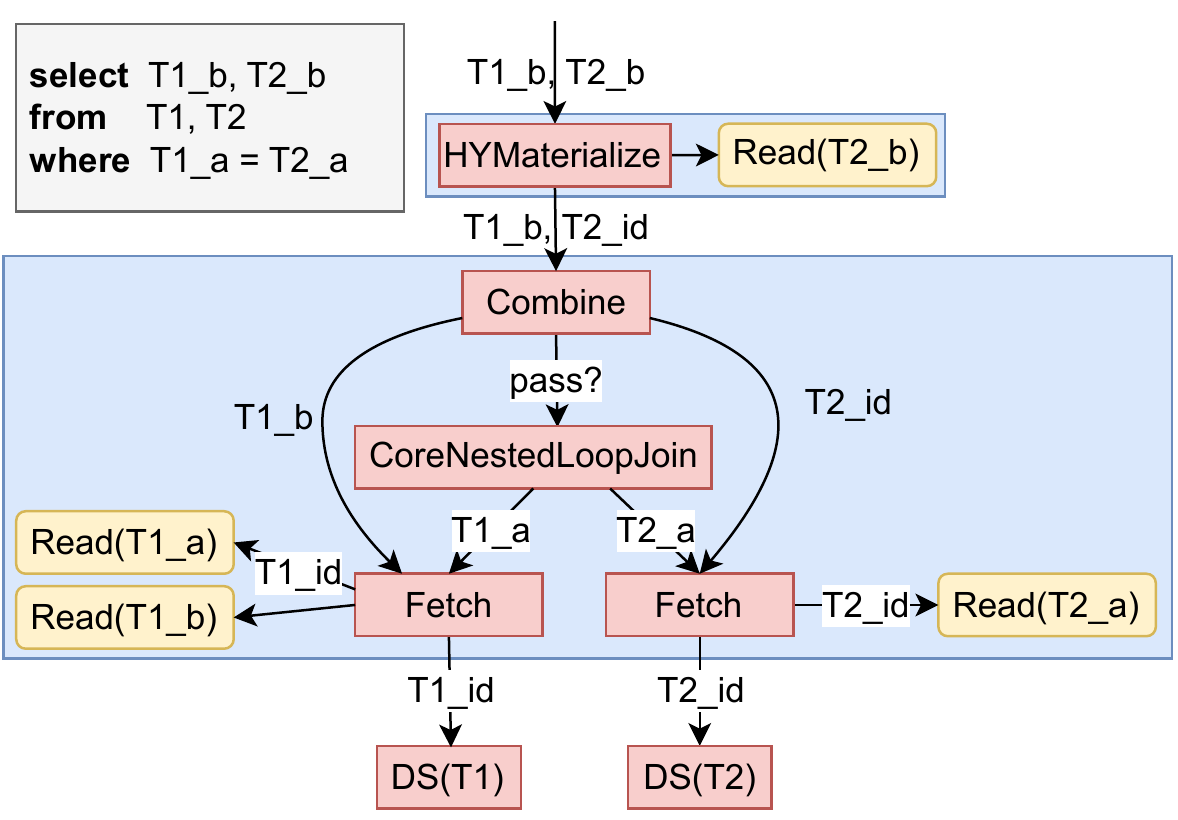}
    \caption{Hybrid materialization}
\end{subfigure}\hfill
\begin{subfigure}{0.4465\linewidth}%
    \centering
    \includegraphics[width=0.97\linewidth]{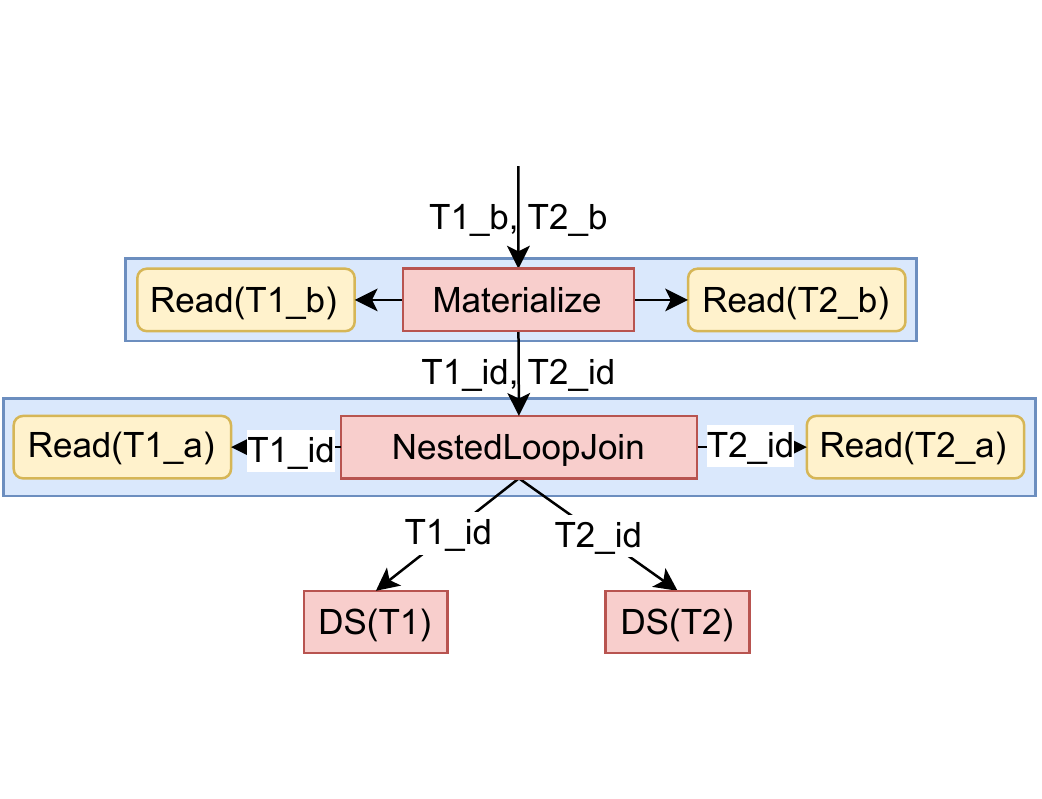}
    \caption{Ultra-late materialization}
\end{subfigure}
\caption{Join between two big tables}
\label{fig:local:ulm}
\end{figure*}

Suppose that we have a query with a nested-loop join between two big tables $T1$ and $T2$, as in Fig.~\ref{fig:local:ulm}. Assume for the sake of simplicity that:

\begin{enumerate}
    \item for each $T2\_a$ value there is a single corresponding (equal) value in $T1\_a$,
    \item nested-loop join preserves the order of records in the right subtree.
\end{enumerate}

Both hybrid and ultra-late materialization processes nested-loop join in a similar manner: for each $T2\_a$ find a corresponding $T1\_a$ and build the next resulting record. The main difference is that hybrid materialization can read $T1\_b$ before join, when $T1$ positions are still sequential. At the same time, ultra-late materialization will force query engine to read $T1\_b$ after the join, in the \lstinline{Materialize} operator.

In the worst case, the nested-loop join in ultra-late materialization results in sequential $T2\_id$s, but their corresponding $T1\_id$s will be sparsely spread across multiple $T1\_b$ pages on disk and they will not be ordered. For example, suppose that the join returned the following position pairs: $\{T2\_id = 1, T1\_id = 5\}$, $\{T2\_id = 2, T1\_id = 1000000\}$, $\{T2\_id = 3, T1\_id = 3000\}$. Such $T1\_id$s require expensive random access to read $T1\_b$s, which is known as the out-of-probing problem~\cite{columns_tutorial}. Hybrid materialization, on the other hand, reads $T1\_b$ before the join, sequentially, and thus efficiently.

A similar problem arises if we swap the nested-loop join to a hash-join. In case of ultra-late materialization hash-join creates a hash-table of $T1\_id$, which later gives ${T2\_id, T1\_id}$ pairs with sequential $T2\_id$s, but randomly distributed $T1\_id$s.

Out-of-order probing may appear in many cases even if the column itself fits into memory. This may happen because DBMS has to spend a part of this memory on hash tables, external sorting, and other concurrently running queries.

\subsection{Local case. Late vs hybrid materialization}

In the local case hybrid materialization has the same advantages over the late materialization as the ultra-late. Specifically, late materialization forces the query engine to:

\begin{enumerate}
    \item read a lot of data before joins even when joins are filtering, i.e. discard most rows;
    \item copy more data during pushing values through subsequent operators,
    \item increase the size of data structures such as hash tables.
\end{enumerate}

Two last points give rise to less explicit, but important drawbacks:
\begin{enumerate}
    \item reduced cache efficiency (locality),
    \item increased number of \lstinline{getNext()} calls,
    \item reduced overall DBMS query bandwidth, e.g., the system will have less memory for hash tables of concurrently running queries.
\end{enumerate}

\subsection{Distributed case. Ultra-late vs hybrid materialization}
\label{sec:adv:ulm}
In case of distributed queries, the cost of repeated access is usually significant even without the out-of-order probing problem. Suppose that executing some operator on Node1 we would like to read $T1\_a$ using $T1$ positions, while $T1\_a$ resides on Node2. In this case, we have (see Section 4.4. in~\cite{DOLAP22-newarch}) to do the following:

\begin{enumerate}
    \item send $T1$ positions back to Node2,
    \item wait until $T1\_a$ is read from disk on Node2,
    \item wait until $T1\_a$ is delivered back from Node2 to Node1.
\end{enumerate}

This process is quite costly as steps 1 and 3 are bounded by network latency and bandwidth. We could try to make them asynchronous, but, to read $T1\_a$, we need to know $T1$ positions which are received from previous operators on Node1. Thus, we cannot easily request $T1\_a$ in advance.

In the case of hybrid materialization, the query engine can send $T1\_a$ together with $T1$ positions to Node1. Thus, $T1$ data is sent in one direction from Node2 directly to Node1 without additional loops. It speeds up query execution and makes it easier to analyze and optimize plans due to simpler data flows.

\subsection{Distributed case. Late vs hybrid materialization}
\label{sec:adv:lm}
When operating on positions, late materialization has the same drawback as the one described above for ultra-late materialization. However, late materialization uses positions only in predicates, which are usually executed within a node. Thus, the cost of repeated accesses for late materialization should be roughly the same in the local and distributed cases.

However, distributed queries with late materialization have another drawback related to pushing extra data through operators. Suppose we have a query that:

\begin{enumerate}
    \item filters records by $T1\_a$ on Node1 that has $T1$ data locally,
    \item joins some tables with $T1$ using several distributed joins, and one of the join predicates uses $T1\_a$;
    \item returns records to the user on Node2 with attributes that include $T1\_b$, $T1\_c$, $T1\_d$.
\end{enumerate}

Using late materialization, we have to read all of the $T1\_a$, $T1\_b$, $T1\_c$, and $T1\_d$ after filters on Node1 and push them through all the distributed operators. Depending on the attribute size, this can increase the amount of data send over the network by several times. Network communication is expensive and is usually a bottleneck in distributed systems, so increasing network traffic by several times will significantly slow down query processing. 

Using hybrid materialization, the query engine can materialize only $T1\_a$ after filters on Node1 and read $T1\_b$, $T1\_c$, $T1\_d$ later, if Node2 has $T1$ data. This way it can decrease network load and speed up query processing. Note that due to joins the late materialization plan has to send $T1\_b$, $T1\_c$, and $T1\_d$ for each result row. Even if $T1$ has a small row cardinality, the number of rows pushed through the plan can be very large. This can happen when other tables participating in joins are large and contain many matching values in join attributes.

\subsection{Wrap-up and Takeouts}

We have shown that hybrid materialization can speed up query processing (compared to late and ultra-late materialization) for both local and distributed cases. The potential performance boost arises from flexibility of the hybrid query model, in which the query engine can materialize individual attributes earlier or later during query execution. Therefore, hybrid materialization can overcome both the out-of-order probing problem of ultra-late materialization and the problem of reading and processing extra data in late materialization.

We believe hybrid materialization deserves implementation and further investigation. It can be especially interesting in a distributed context, since distributed systems are susceptible to slow network due to the amount of data they send over it. We have pointed out two cases when distributed queries can benefit from hybrid materialization, but there should be more, and they could potentially be associated with distributed join algorithms.

\section{Experimental evaluation}
\label{sec:experiments}
As we have previously noted, hybrid materialization can be useful for both distributed and centralized cases. We decided to concentrate on centralized experiments in this paper as a first step. A lot of systems support only local query execution, which is also a more stable case, independent of such implementation details as network subsystem optimizations, network configuration, network latency and bandwidth. For the distributed cases, we have discussed potential advantages of hybrid materialization in Sections~\ref{sec:adv:ulm} and~\ref{sec:adv:lm}. Detailed distributed experiments are a good subject for a separate future study.

Our investigation is based on the TPC-H benchmark~\cite{TPCH} with the Scale Factor (SF) in a [1, 50] range. Table~\ref{tbl:tpch_tables} shows the precise number of rows in individual TPC-H tables for easier experiment analysis.

We have replaced the previously used SSB~\cite{DOLAP22-newarch, SSB} with the TPC-H as its queries have more complicated dataflows, while still being fairly simple to allow for engine comparison instead of optimizer output.

Specifically, we start with a simplified version of Q5, which we use to demonstrate query plans for various materialization strategies and compare their performance. Then, we continue with a more realistic (and complex) query based on the slightly modified SPJ part of Q9 with multiple joins between various data tables.

\begin{table}
\centering
\caption{TPC-H table size}
\begin{tabular}{ll}

Table name & Num tuples \\ \hline \hline
region & 5 \\ \hline
nation & 25 \\ \hline
supplier & SF * 10 000 \\ \hline
customer & SF * 150 000 \\ \hline
part & SF * 200 000 \\ \hline
partsupp & SF * 800 000 \\ \hline
orders & SF * 1 500 000 \\ \hline
lineitem & SF * 6 000 000 \\ \hline

\end{tabular}
\label{tbl:tpch_tables}
\end{table}

All experiments were run on a desktop with the following hardware: AMD Ryzen 9 3900X, GIGABYTE X570 AORUS ELITE, Kingston HyperX FURY Black HX434C16FB3K2/32 32GB, 512 GB SSD M.2 Patriot Viper VPN100-512GM28H. The following software versions were used: Ubuntu 22.04 LTS, GCC 11.3.0. Each experiment was run without warm-up and the OS page cache was dropped before each launch. Since each run yielded almost the same numbers, we present results that were averaged over ten launches.

\subsection{Analytic workloads}

Hybrid materialization can co-exist with late and ultra-late materialization as well as emulate both of them with special \lstinline{Fetch-Combine} pairs, as described in Section~\ref{sec:hybrid:internals}.

This means that if ultra-late materialization is more than two times faster than late materialization on SSB~\cite{DOLAP22-newarch}, hybrid materialization will perform close to the ultra-late. The purpose of the paper is to show cases when both late and ultra-late strategies have fundamental limitations which arise from their flow, not from implementation details. Therefore, we have decided to move away from the SSB which is rather limited in terms of possible queries and is ill-fit for our purposes. The issue with SSB is that it considers a particular case with a single large fact table and several very small dimension tables.

\subsection{Hybrid materialization strategy}

In the hybrid query model, there is a tension between the cost of materializing an attribute early and the cost of pushing positions through subsequent plan operators with materializing this attribute later. For some attributes of a particular query, reading data early is an efficient way to prevent out-of-order probing and costly remote access. However, for other attributes of the same query we can materialize data later, decreasing the size of in-memory data structures and the cost of pushing data blocks through operators.

Building a cost model and a full-fledged optimizer for hybrid materialization is an interesting topic. Such optimizer should at least account for materialization point (time) of each individual attribute, size of various tables, and amount of available RAM on nodes. However, the purpose of this paper is to show how column-store can support hybrid materialization and prove that there are cases when hybrid query model is quite beneficial and therefore deserves further investigation. 

\begin{figure}[htb]
    \centering
    \includegraphics[width=\linewidth]{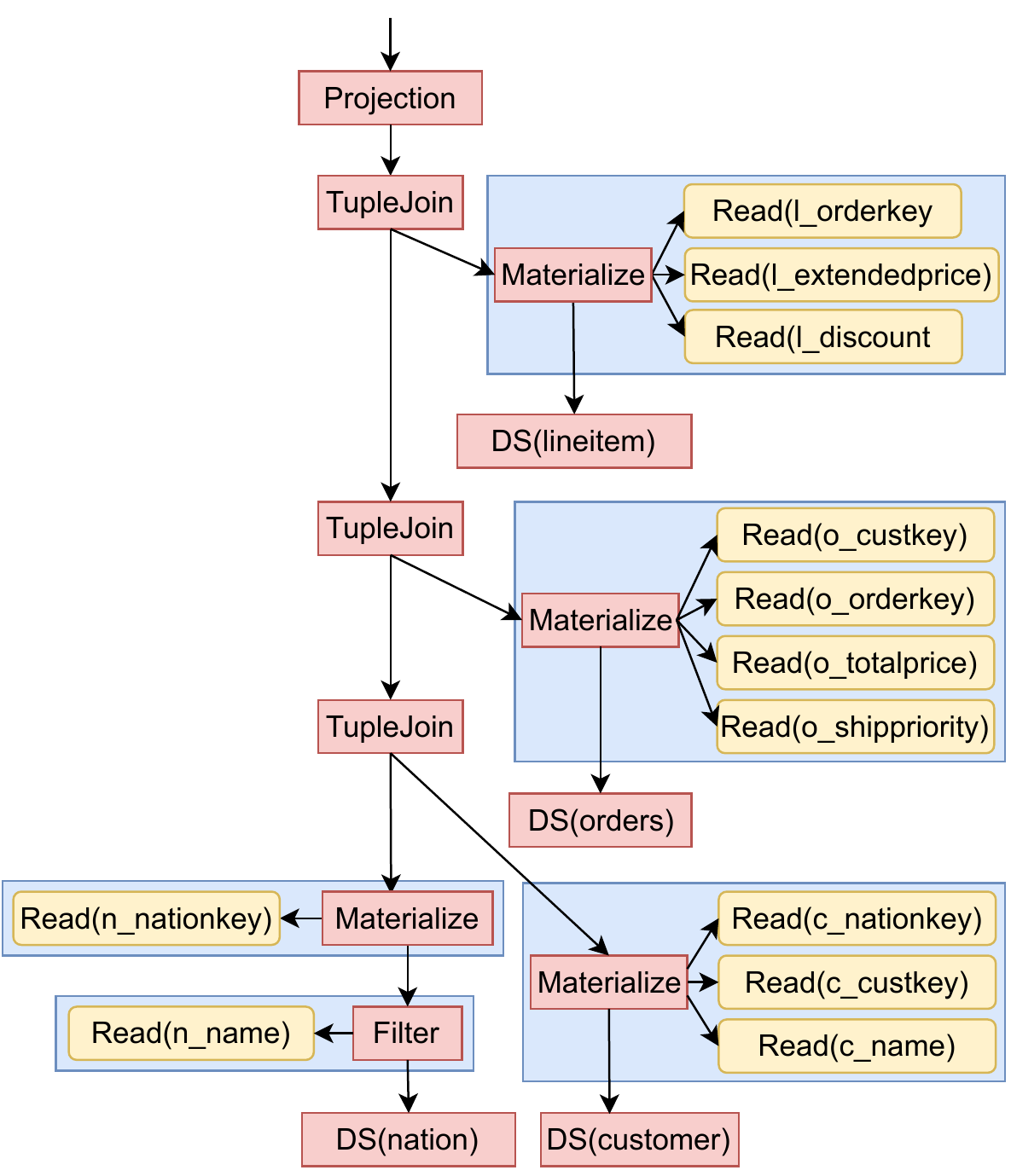}
    \caption{Late materialization in PosDB'22}
    \label{fig:q5_lm}
\end{figure}

Thus, instead of building a query optimizer, we use a simple heuristic, which is as follows: we materialize data as late as reasonably possible. This means that we create hybrid materialization plans similarly to ultra-late materialization and then materialize some attributes earlier, aiming to mitigate the issues of out-of-order probing and repeated access.

\subsection{Experiment 1}

In the first experiment, we are using a modified SPJ part of the TPC-H Q5. We have intentionally omitted aggregation and subquery parts since materialization strategy mostly affects joins and predicates. We have also added a join with a filtered \lstinline{nation} table and a selection of the \lstinline{c_name} column:

\begin{lstlisting}[language=sql,label={listing:q5_query},basicstyle=\small]
SELECT
    c_name, o_totalprice, o_shippriority, l_orderkey,
    l_extendedprice * (1 - l_discount)
FROM 
    nation, customer, orders, lineitem
WHERE
    n_name = ALGERIA AND
    n_nationkey = c_nationkey AND
    c_custkey = o_custkey AND
    o_orderkey = l_orderkey;
\end{lstlisting}

We have designed this query on the following principles:

\begin{enumerate}
    \item Three joins between three medium-large tables.
    \item There is enough RAM to use hash joins.
    \item Fourth join with a fixed-size table, \lstinline{nation}, to make one of the joins ``filtering'' with $1/25$ selectivity (the ``ALGERIA'' filter). It makes late materialization expensive, as it should process all the records of \lstinline{orders} and \lstinline{lineitem} attributes.
    \item Chain of joins makes \lstinline{customer} positions (\lstinline{c_id}s) unordered by the moment when \lstinline{c_name} values are read within ultra-late materialization. It makes ultra-late materialization expensive due to out-of-order probing.
\end{enumerate}

\begin{figure}[htb]
    \centering
    \includegraphics[width=0.874\linewidth]{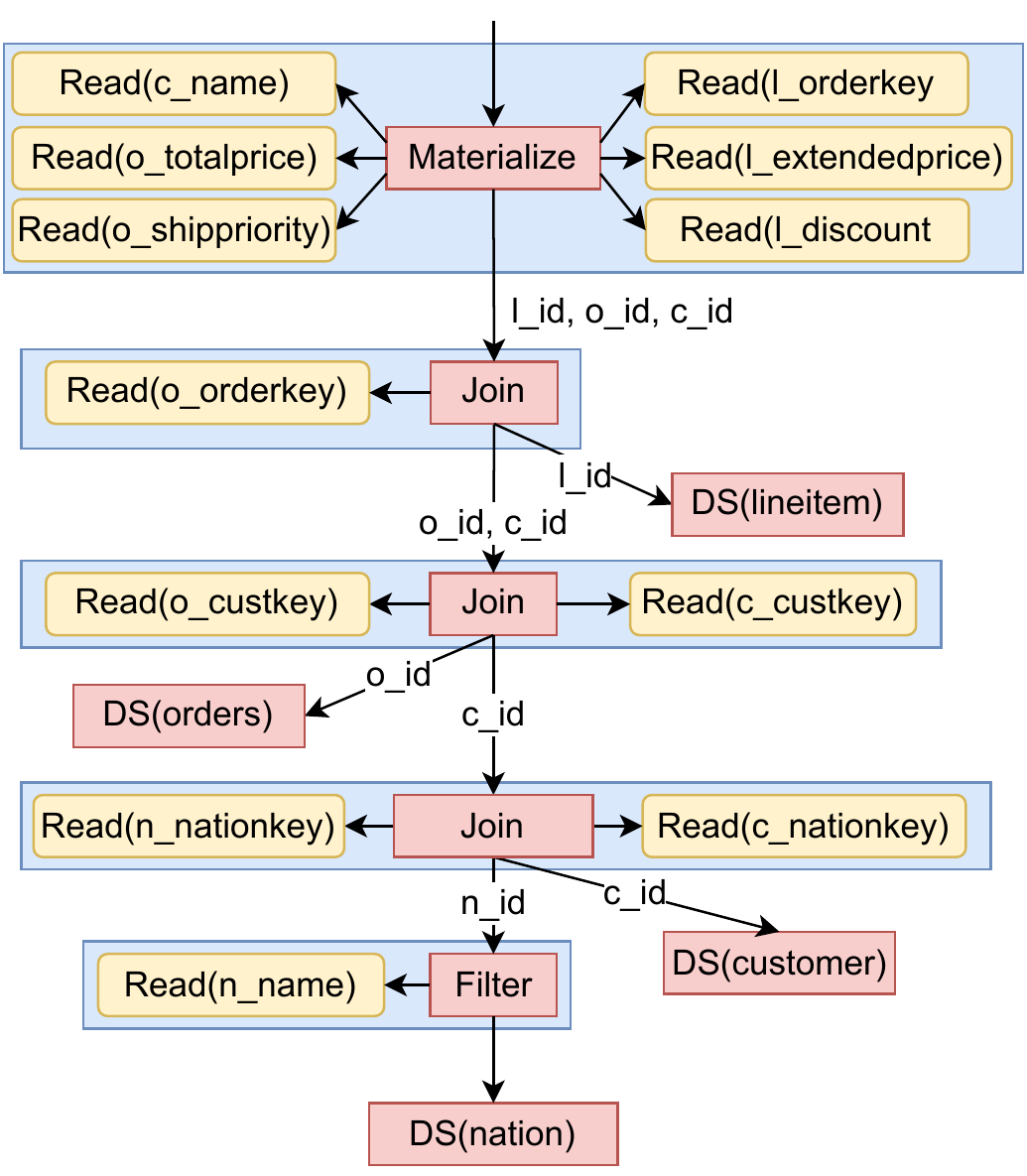}
    \caption{Ultra-late materialization in PosDB'22}
    \label{fig:q5_ulm}
\end{figure}

Fig.~\ref{fig:q5_lm} and Fig.~\ref{fig:q5_ulm} demonstrate specific query plans in late and ultra-late materialization models, which are what PosDB'22 uses.

\begin{figure}[htb]
    \centering
    \includegraphics[width=0.95\linewidth]{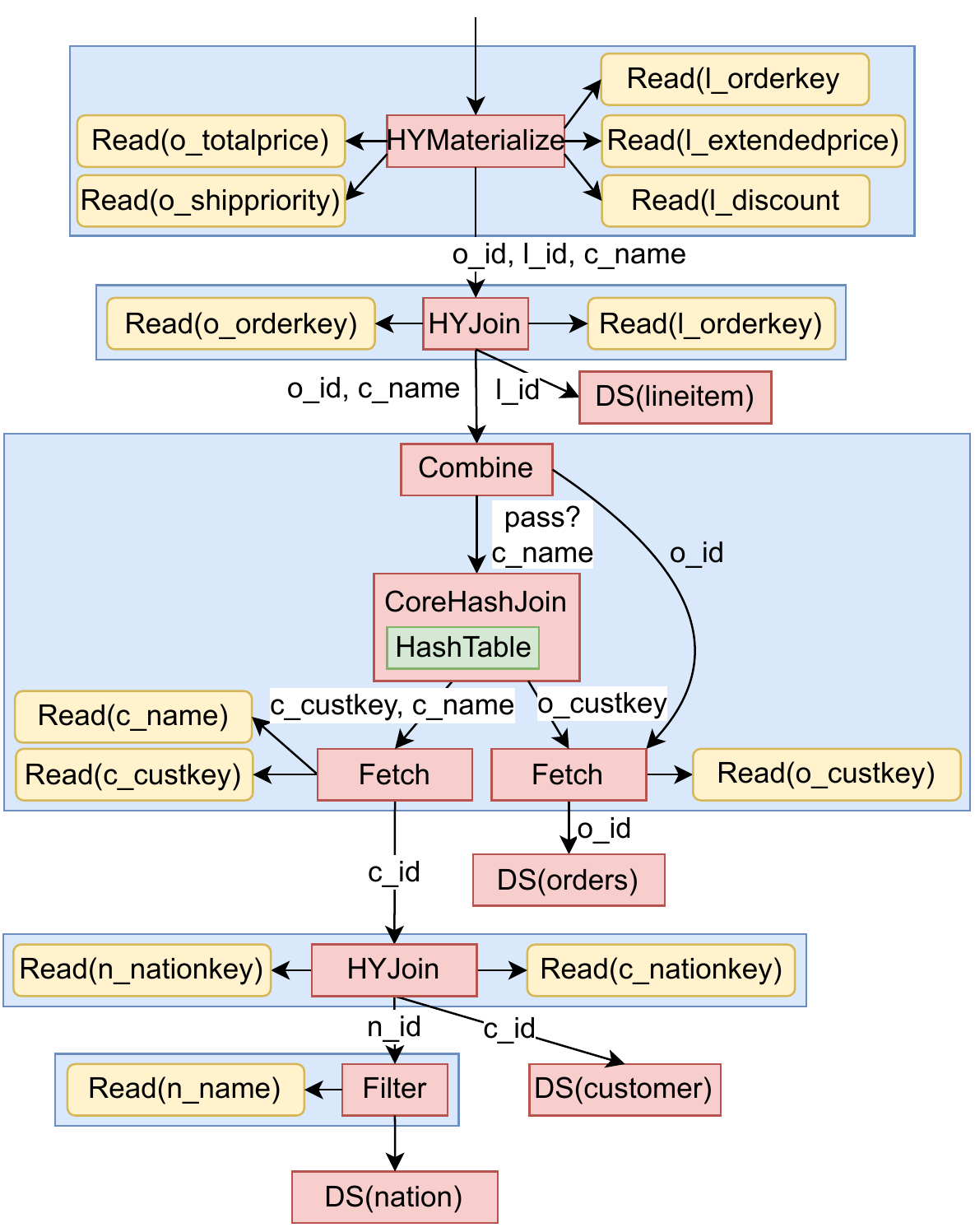}
    \caption{Hybrid materialization in PosDB'23}
    \label{fig:q5_hm}
\end{figure}

Hybrid materialization can combine strengths of both approaches, materializing \lstinline{orders} and \lstinline{lineitem} data in the final \lstinline{HYMaterialize}, but caching \lstinline{c_name} when \lstinline{customer} positions are still ordered. Thus, we reduce the amount of processed data and avoid out-of-order probing. Fig.~\ref{fig:q5_hm} shows query plan with hybrid materialization in PosDB'23.

\begin{figure*}[!t]
\begin{subfigure}{0.49\linewidth}%
    \centering
    \includegraphics[width=\linewidth]{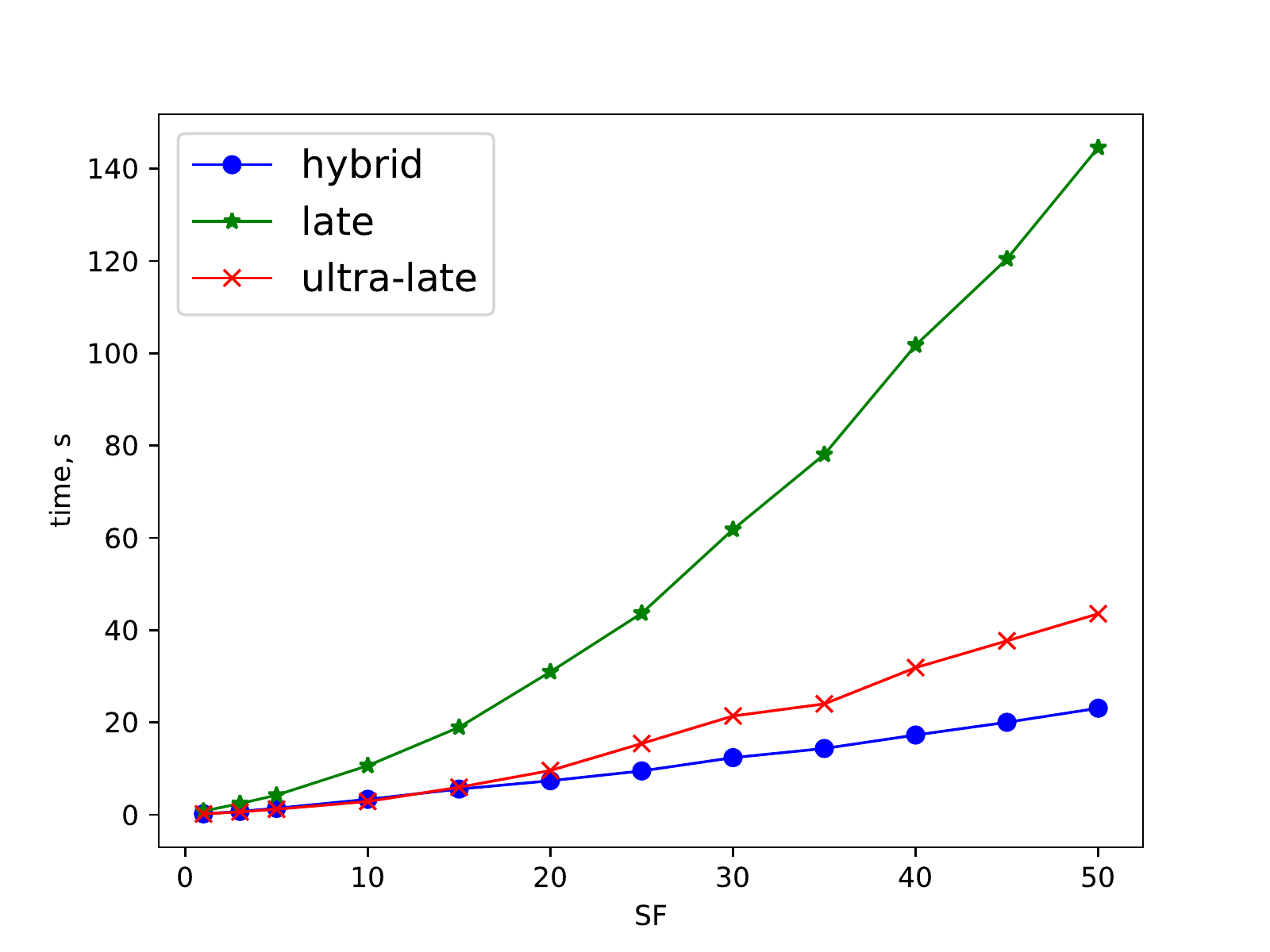}
    \caption{With out-of-order probing}
    \label{fig:q5_test}
\end{subfigure}\hfill
\begin{subfigure}{0.49\linewidth}%
    \centering
    \includegraphics[width=\linewidth]{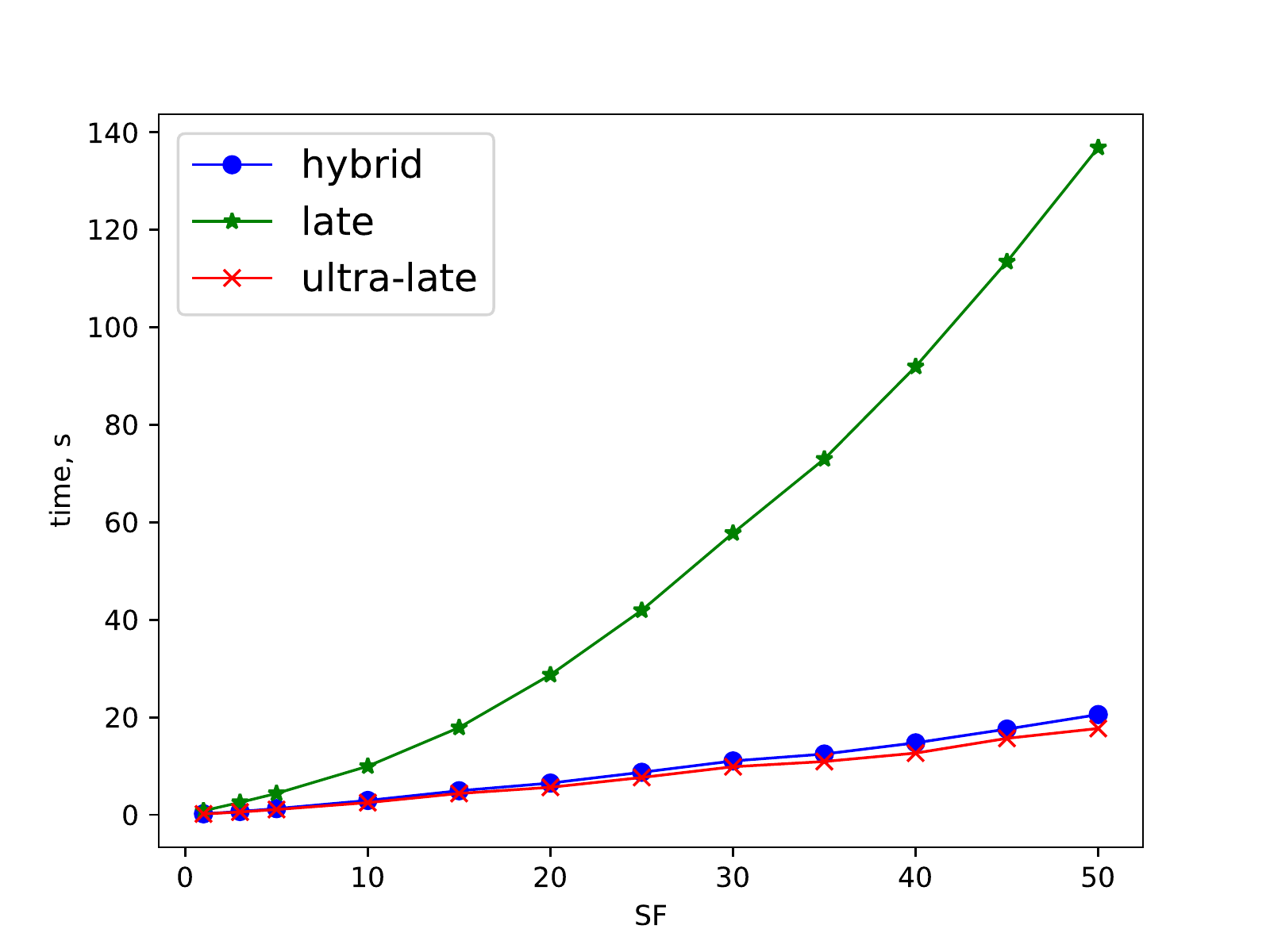}
    \caption{Without out-of-order probing}
    \label{fig:q5_test_fixed}
\end{subfigure}

\caption{Modified Q5 of TPC-H}
\end{figure*}

To prove our points, we first evaluated all the strategies on TPC-H with SF ranging from 1 to 50, Fig.~\ref{fig:q5_test}. It can be seen that on small SF ultra-late and hybrid materialization have roughly the same performance and are significantly faster than late materialization. On bigger scale factors, the out-of-order probing problem becomes more evident as the whole \lstinline{c_name} column cannot be easily put into RAM together with all the hashtables. Thus, ultra-late materialization becomes more and more expensive. On the contrary, hybrid materialization demonstrates moderate  growth and finally becomes almost two times faster than ultra-late on SF 50.

We repeat the same experiment once again, this time without \lstinline{c_name} in the \lstinline{select} clause of the query. It removes out-of-order probing and as we can see in Fig.~\ref{fig:q5_test_fixed}, both hybrid and ultra-late materialization are faster than late materialization on all SFs. Hybrid materialization is several percent slower than ultra-late due to implementation overhead.

\subsection{Experiment 2}

In the second experiment, we are using the SPJ part from TPC-H Q9, which gives us a more realistic query example with a more complicated join structure. The subquery and aggregation parts are again omitted in order to concentrate on joins and predicates:

\begin{lstlisting}[language=sql,label={listing:q9_query},basicstyle=\small]
SELECT
    n_name, o_totalprice, l_extendedprice * (1 - l_discount) -
    - ps_supplycost * l_quantity
FROM
    part, supplier, lineitem, partsupp, orders, nation
WHERE
    s_suppkey = l_suppkey
    AND ps_suppkey = l_suppkey
    AND ps_partkey = l_partkey
    AND p_partkey = l_partkey
    AND o_orderkey = l_orderkey
    AND s_nationkey = n_nationkey
    AND p_size > 25;
\end{lstlisting}

We replaced \lstinline{extract(year from o_orderdate)} with \lstinline{o_totalprice} and \lstinline{p_name like '\%:?\%'} with \lstinline{p_size > 25} for simplicity.

\begin{figure}[htb]
    \centering
    \includegraphics[width=\linewidth]{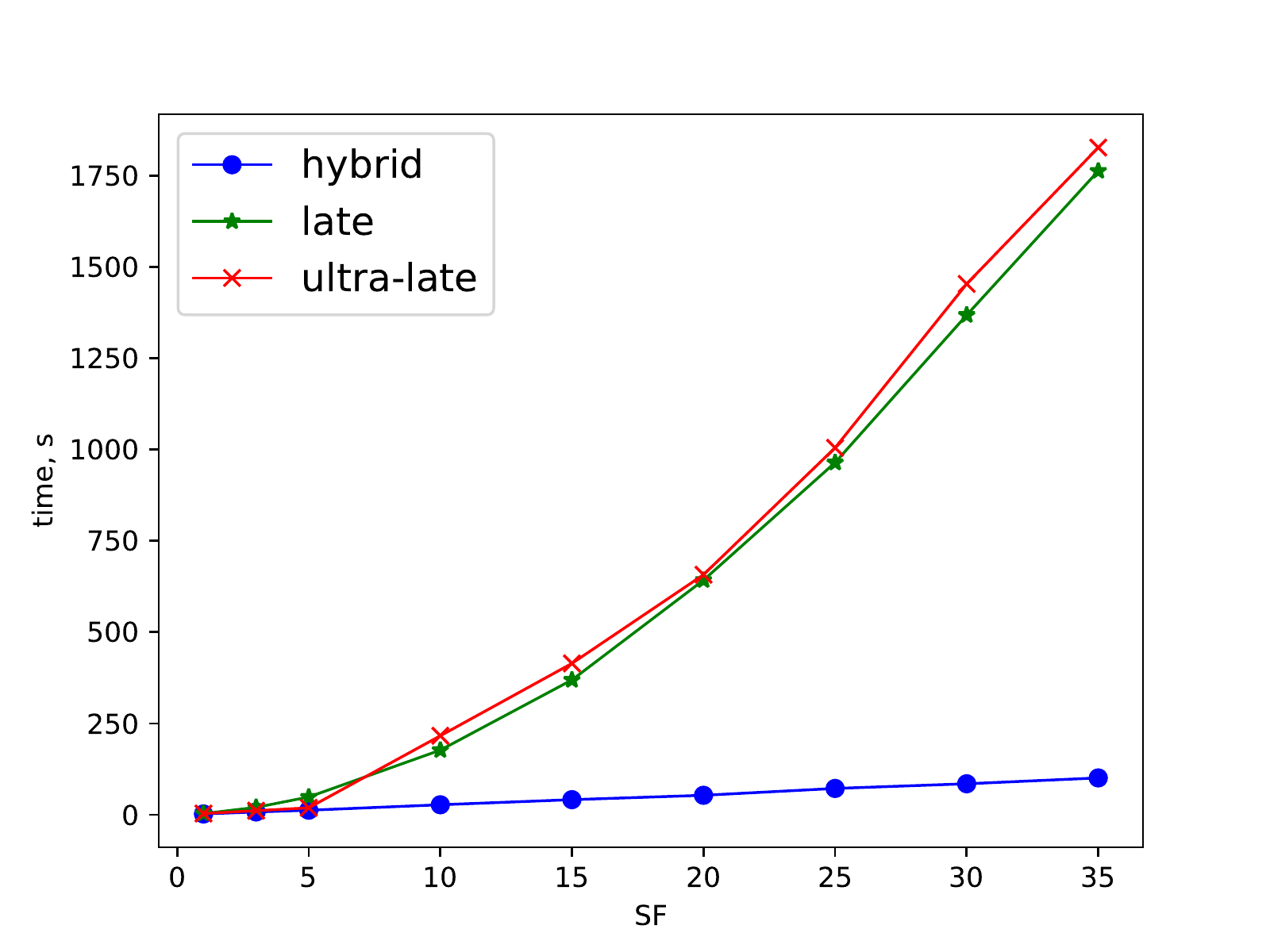}
    \caption{Modified TPC-H Q9}
    \label{fig:q9_test}
\end{figure}

We studied the performance of this query under all the materialization strategies on TPC-H with SF ranging from 1 to 35, as shown in Fig.~\ref{fig:q9_test}. It can be seen that hybrid materialization performs the best, with more than twofold speedup. Late materialization is slower because the \lstinline{p_size} predicate together with joins between \lstinline{part}, \lstinline{partsupp} and \lstinline{lineitem} filters a lot of records similarly to a filtering join. The ultra-late strategy is slower because it has to read \lstinline{ps_supplycost} with random access (i.e., perform out-of-order probing).

\subsection{Summary}

In this study, we concentrated on centralized experiments based on the TPC-H benchmark and compared various materialization strategies. Our findings show that:

\begin{enumerate}
    \item Ultra-late materialization becomes slow when out-of-order access is necessary.
    \item Late materialization becomes slow when a query plan contains filtering joins.
    \item Hybrid materialization combines both approaches and can provide almost twofold speedup.
\end{enumerate}

\section{Future plans}
\label{sec:future}

We plan to extend our study of hybrid materialization in several directions:

\begin{enumerate}
    \item Comparing hybrid materialization and late materialization with specific optimizations such as sideways information passing~\cite{vertica}.
    \item Digging deeper into distributed cases, finding more dataflow patterns that can benefit from hybrid materialization and  studying them experimentally.
    \item Developing an optimal strategy for choosing the materialization point of each attribute in the hybrid query model.
\end{enumerate}

\section{Conclusion}
\label{sec:conclusion}

In this paper, we continued our previous investigation of materialization strategies in disk-based column-stores~\cite{DOLAP22-newarch}. First, we reviewed existing early, late, and ultra-late materialization strategies and their main aspects. We showed that all the strategies lack flexibility, and there are cases when each strategy runs into fundamental limitations.

Following this, we proposed a hybrid materialization strategy which can use strengths of all the existing strategies, both delaying data materialization and avoiding expensive repeated data access. Hybrid position-value processing was shortly mentioned in existing research papers, but hybrid materialization has not been extensively studied.

Thus, we have presented a hybrid block structure, hybrid operator internals, and the overall algebra of hybrid operators. Then, we showed four local and distributed cases when hybrid strategy has fundamental advantages over late and ultra-late materialization. Finally, we implemented our approach inside PosDB~--- a distributed, disk-based column-store.

The experimental evaluation using modified versions of Q5 and Q9 from the TPC-H~\cite{TPCH} benchmark demonstrated almost twofold speedup due to hybrid materialization.

\section*{Acknowledgments}
We would like to thank Anna Smirnova for her help with the preparation of the paper. 

\bibliographystyle{IEEEtran}

\bibliography{FRUCTexample}



\end{document}